\documentclass[fleqn,usenatbib]{mnras}

\usepackage{newtxtext,newtxmath}

\usepackage[T1]{fontenc}

\DeclareRobustCommand{\VAN}[3]{#2}
\let\VANthebibliography\thebibliography
\def\thebibliography{\DeclareRobustCommand{\VAN}[3]{##3}\VANthebibliography}

\usepackage{graphicx}	
\usepackage{xcolor}
\usepackage{amsmath}	
\usepackage{breqn}

\newcommand{\ko}{\textbf{K$_{0}$}\,}
\newcommand{\Beq}{\textbf{B$_{\rm eq}$}\,}
\newcommand{\Inulow}{I$_{\nu, \, \nu\, = \, 0.33 \rm{GHz}}$ \,}
\newcommand{\Inuhi}{I$_{\nu, \, \nu\, = \, 4.83 \rm{GHz}}$\,}
\newcommand{\Inu}{I$_{\nu}$\,}

\title[Synchrotron Emission on FIRE]{Synchrotron Emission on FIRE: Equipartition Estimators of Magnetic Fields in Simulated Galaxies with Spectrally-Resolved Cosmic Rays}

\author[S. Ponnada et al.]{
\vspace{0.1cm}
\parbox[t]{\textwidth}{Sam B. Ponnada,$^{1}$\thanks{E-mail: sponnada@astro.caltech.edu}
Georgia V. Panopoulou,$^{2}$
Iryna S. Butsky,$^{1}$
Philip F. Hopkins,$^{1}$
Raphael Skalidis,$^{1}$
Cameron Hummels,$^{1}$
Eliot Quataert,$^{3}$
Du\v{s}an Kere\v{s},$^{4}$
Claude-Andr\'e Faucher-Gigu\`ere,$^{5}$
Kung-Yi Su$^{6}$}
\\
$^{1}$California Institute of Technology, TAPIR, Mailcode 350-17, Pasadena, CA 91125, USA\\
$^{2}$Department of Space, Earth and Environment, Chalmers University of Technology, 412 93, G\"oteborg, Sweden\\
$^{3}$ Department of Astrophysical Sciences, Princeton University, Princeton, NJ 08544, USA \\
$^{4}$ 
Department of Physics, Center for Astrophysics and Space Sciences, University of California San Diego
, 9500 Gilman Drive, La Jolla, CA 92093, USA\\
$^{5}$Department of Physics and Astronomy and CIERA, Northwestern University, 2145 Sheridan Road, Evanston, IL 60208, USA\\
$^{6}$Black Hole Initiative, Harvard University, 20 Garden Street, Cambridge, MA 02138, USA\\
}

\date{Accepted 2023 December 21. Received 2023 December 14; in original form 2023 September 7}

\pubyear{2023}

\begin{document}
\label{firstpage}
\pagerange{\pageref{firstpage}--\pageref{lastpage}}
\maketitle

\begin{abstract}
Synchrotron emission is one of few observable tracers of galactic magnetic fields (\textbf{B}) and cosmic rays (CRs). Much of our understanding of \textbf{B} in galaxies comes from utilizing synchrotron observations in conjunction with several simplifying assumptions of equipartition models, however it remains unclear how well these assumptions hold, and what \textbf{B} these estimates physically represent. Using FIRE simulations which self consistently evolve CR proton, electron, and positron spectra from MeV to TeV energies, we present the first synthetic synchrotron emission predictions from simulated L$_{*}$ galaxies with "live" spectrally-resolved CR-MHD. We find that synchrotron emission can be dominated by relatively cool and dense gas, resulting in equipartition estimates of \textbf{B} with fiducial assumptions underestimating the "true" \textbf{B} in the gas that contributes the most emission by factors of 2-3 due to small volume filling factors. Motivated by our results, we present an analytic framework that expands upon equipartition models for estimating \textbf{B} in a multi-phase medium. Comparing our spectrally-resolved synchrotron predictions to simpler spectral assumptions used in galaxy simulations with CRs, we find that spectral evolution can be crucial for accurate synchrotron calculations towards galactic centers, where loss terms are large.
\end{abstract}

\begin{keywords}
ISM: cosmic rays -- ISM: magnetic fields -- galaxies: formation -- methods: numerical -- methods: analytical
\end{keywords}



\section{Introduction}
Magnetic fields (\textbf{B}) and relativistic charged particles (cosmic rays, hereafter CRs) are of significant importance in astrophysics. Both are known to provide a significant source of non-thermal pressure support in the interstellar medium (ISM) \citep{Boulares1990}, influence the physics of giant molecular clouds \citep[for relevant reviews, see][]{Crutcher2011,beck_synthesizing_2020}, and magnetic fields determine the transport of the dynamically coupled CRs through the ISM and into the CGM \citep{zweibel_microphysics_2013}, therefore influencing the structure of galactic outflows \citep[see][for relevant reviews]{zhang_review_2018,owen_cosmic_2023,ruszkowski_cosmic_2023}. 

A major difficulty in our understanding of galactic magnetic fields, and subsequently on our understanding of CR propagation and ensuing effects, is the measurement of magnetic field strengths and geometries. A unique extragalactic probe of \textbf{B} and CRs comes from the synchrotron emission radiated by CRs as they gyrate around magnetic field lines \citep{Ginzburg1965}. Most estimates of magnetic field strengths in dwarf and spiral galaxies \citep{fitt_magnetic_1993,beck_magnetic_2000,Chyzy2011,Fletcher2011,basu_magnetic_2013,beck_magnetic_2015} are derived indirectly from the intensity of this non-thermal synchrotron emission, often at radio frequencies, with the key assumption of energy equipartition between CR protons and \textbf{B}, used to resolve their formal strict degeneracy. 

As discussed in \citet{beck_revised_2005,stepanov_observational_2014,seta_revisiting_2019}, the energy equipartition method is motivated by the intertwined dynamics of cosmic rays and magnetic fields, in approximate pressure equilibrium in the local, warm ISM. While this technique has been applied to several radio observations of galaxies across a wide range of spatial scales \citep{beck_synthesizing_2020}, it is not clear whether \textbf{B} and CRs are in approximate pressure equilibrium at all spatial scales, and across the large dynamic range of ISM gas densities and temperatures. Additionally, the method is subject to further assumptions regarding the spectrum, the spatial distribution of CR electrons (CRe) and protons (CRp), as well as the effective emitting volume, which are potentially order-of-magnitude uncertain given the multi-phase nature of the ISM.

Given these assumptions and caveats, it is unclear how to interpret the equipartition estimates of \textbf{B}. \textbf{B} is known to vary by orders of magnitude with gas density and with ISM phase \citep{Crutcher2011, han_observing_2017}, and while equipartition-inferred estimates of \textbf{B} may inform us about some volumetric and LOS-averaged estimator of \textbf{B}, it remains unclear how this maps to measuring \textbf{B} \textit{within} a given ISM phase or gas density. In other words, what phase of the ISM dominates the synchrotron emission, and how well does the equipartition estimate of \textbf{B} trace it?

Exploring questions about synchrotron emission and equipartition in numerical simulations has been limited in scope by the computational complexity of self-consistently evolving \textbf{B} and CRs, in the context of the multi-phase ISM, which exhibits great spatial and temporal variation in gas properties across cosmological time. Simulations have only recently been able to evolve magnetic fields \citep{peng_magnetohydrodynamic_2009,pakmor_magnetic_2014,Marinacci2014, Su2017,Rieder2017,Butsky2017, Martin-Alvarez2018, Ntormousi2020} including CRs as a coupled fluid term \citep{booth_simulations_2013,girichidis_launching_2016, Butsky2018,chan_cosmic_2019, buck_effects_2020,Werhahn2021,werhahn_cosmic_2021b,werhahn_cosmic_2021, Hopkins2020,farcy_radiation-magnetohydrodynamics_2022}, though it varies whether the simulations in these studies are cosmological, in addition to differences in resolution, physics prescriptions, and numerical methods. Furthermore, these simulations which incorporate a fluid treatment of CRs often take the so-called "single-bin" approximation, evolving solely the $\sim$1-10 GeV CR proton energy density (or a constant underlying spectrum, with transport coefficients corresponding to the 1-10 GeV CR protons) and by construction fail to capture the CRe spectra, which are required for predictive synchrotron calculations.

Recent algorithmic advances have pushed this boundary of explicitly modeling the CR spectrum \citep{yang_spatially_2017,girichidis_spectrally_2020,ogrodnik_implementation_2021}, allowing us to evolve the spectra of various CR species in live-kinetic magneto-hydrodynamic (MHD) simulations of galaxy formation \citep{hopkins_first_2022}. Additionally, detailed study of \textbf{B} within this family of simulations (which share the same physical prescriptions, up to the novel spectral treatment of CRs) has shown that they produce realistic \textbf{B} strengths and geometries in resolved ISM phases \citep{Ponnada2022}. 

In this work, we forward-model synchrotron emission from high-resolution simulations of galaxy formation with cosmological initial conditions. We present our methodology for computing synchrotron emission from a set of L$_*$ galaxies from the Feedback in Realistic Environments project (FIRE)\footnote{\url{https://fire.northwestern.edu/}} simulation suite \citep{Hopkins2018,hopkins_fire-3_2022} in Section \ref{sec:methods}. In Section \ref{sec:results}, we compare our results for the spectrally-resolved FIRE runs to relevant observations, and explore how equipartition estimates of magnetic field trace the underlying magnetic fields within these simulations. Through resolving multi-phase ISM structure in our simulations, we disentangle how different phases of the ISM contribute to the resulting synchrotron intensity, and physically interpret equipartition measurements of \textbf{B}, discussing how well each of the fundamental assumptions in empirical equipartition models does or does not apply. In Section \ref{sec:toy_model}, we present a toy model for understanding equipartition magnetic field estimates in the context of a multi-phase ISM motivated by our results in the prior section. Finally, we discuss our results and conclusions and summarize our findings in Section \ref{sec:discussion}.

\section{Methods}\label{sec:methods}

In this section, we will discuss the details of the simulations used in this study, and describe our methodology to compute synchrotron emissivities from the simulations in post-processing.

We investigate three zoom-in simulations of galaxies roughly similar in mass and size to the Milky Way, named \texttt{m12i}, \texttt{m12f}, and \texttt{m12m}. These simulations were shown in previous works to all produce CR spectra consistent with those local ISM (LISM) constraints \citep{hopkins_first_2022}.

All three simulations here have been presented and extensively studied in \citet{hopkins_first_2022} (which did not, however, model their synchrotron properties), to validate that many other properties (e.g. stellar and gas masses, sizes, kinematics, etc) are plausibly consistent with observed galaxies of similar masses. The simulations all have a Lagrangian mass resolution of 56000 M$_{\odot}$ for gas cells, and the typical spatial resolution ranges from $\sim 1-10$\,pc in dense gas.

\subsection{Simulations}\label{sec:sims}

The simulations studied here are all fully-dynamical, cosmological, magnetohydrodynamic-radiation-thermochemical-cosmic ray-gravitational star and galaxy formation simulations. This means they self-consistently follow galaxy formation from cosmological initial conditions at redshifts $>100$ including both dark matter and baryons (in gas and stars), with magnetic fields grown self-consistently from arbitrarily small trace seed fields at $z \approx 100$, with phase structure and thermo-chemistry in the galaxy emerging from cooling with temperatures $T \sim 1-10^{10}\,$K and self-gravity, fully-coupled to multi-band (EUV/FUV/NUV/OIR/FIR) radiation transport, with star formation in the most dense gas ($\gtrsim 1000\,{\rm cm^{-3}}$), and those stars influencing the medium in turn via their injection of radiation, stellar mass-loss, and both Type Ia and core-collapse SNe explosions (followed self-consistently according to standard stellar evolution models). The cosmic ray physics is itself coupled directly to the dynamics, with CRs propagating along magnetic field lines according to the fully general CR transport equations, and interacting with the gas via scattering, Lorentz forces, and losses. 

This means that when we model CRs, all quantities needed to compute the fully non-equilibrium CR dynamics and losses are captured in-code, except for the microphysical CR scattering rate $\nu$. This arises physically from CR pitch-angle scattering off magnetic field fluctuations on gyro-resonant scales -- far smaller than we can possibly resolve ($\sim 0.1\,$au in the warm ISM, for $\sim$\,GeV CRs). As such, the simulations must insert some assumed ``sub-grid'' scattering rate. We stress that the more familiar CR ``diffusion coefficient'' and/or ``streaming speed'' arise self-consistently from $\nu$ in the appropriate limits of the CR dynamics equations (for example, if the CR distribution function is sufficiently close to isotropic and in local flux steady-state, the effective parallel diffusion coefficient is simply $\kappa_{\|}\sim \mathrm{v}_{\rm cr}^{2}/3\nu$).

As mentioned earlier, the simulations presented in Section \ref{sec:results} are the same as those presented in \cite{hopkins_first_2022}, and so we refer the reader for further details therein. Here, we summarize the most pertinent details. These simulations are run with GIZMO\footnote{GIZMO is publicly available at \url{http://www.tapir.caltech.edu/~phopkins/Site/GIZMO.html}.}, in the mesh-free finite-mass mode. All simulations include MHD as treated in \citep{hopkins_accurate_2016,hopkins_constrained-gradient_2016}, and fully-anisotropic Spitzer-Braginskii conduction and viscosity \citep{hopkins_anisotropic_2017,Su2017}. These simulations were restarted at $z \approx$ 0.05 from a high-resolution, cosmological, FIRE-2 `single-bin' CR-MHD simulation of the same galaxy, using the self-consistently evolved CR energy densities in each cell to populate the CR distribution function, and then run for $\sim$500 Myr. Starting from this low redshift and evolving for this runtime are sufficient for the galaxies CR spectra to reach quasi steady-state behavior in the disk and inner CGM at $z=0$. This is true also of secondary leptons, which reach a steady-state on their dominant loss timescale (Coulomb/ionization, diffusive, and radiative losses for $\sim$MeV, $\sim$GeV, and $\gtrsim$ 50 GeV leptons respectively), which never exceed $\sim$10 Myr in the disk and inner halo, as shown in numerical tests of \citep{hopkins_first_2022}. The details of prescriptions for star formation, stellar feedback, CR-MHD spectral evolution, and coupling to gas follow those of \citet{hopkins_fire-3_2022}.

Our implementation of the CR physics is described in detail in \citep{hopkins_first_2022}, and self-consistently includes adiabatic, diffusive re-acceleration, gyro-resonant loss, Coulomb, ionization, hadronic and other collisional, radioactive decay, annihilation, Bremstrahhlung, inverse Compton (IC), and synchrotron loss terms. Hadronic losses are assumed to be dominated by the proton-proton interaction, with total pion loss rates following \citet{Mannheim1994,Guo2008} and those of \citet{Evoli2017} for antimatter. CRs are injected via a power-law spectrum in momentum at SNe (Types Ia \& II) and stellar winds (OB/WR) to neighboring gas cells with fixed fractions $\epsilon^{\rm inj}_{\rm CR}$ = 0.1 and $\epsilon^{\rm inj}_{\rm e}$ = 0.002 of the initial ejecta kinetic energy going into CRs (protons) and leptons. These injection fractions are well motivated by theoretical work on nonlinear diffusive shock acceleration and inverse modeling of observed CR spectra \citep{Caprioli2012, Yuan2012}. Most notably, the fiducial simulations here differ in their treatment of CRs from those presented and analyzed in \citet{Hopkins2020} and \citet{Ponnada2022} by explicitly evolving each bin of the injected CR spectra, following the method of \citet{girichidis_spectrally_2020} and \citet{ogrodnik_implementation_2021} as presented in \citet{hopkins_first_2022}. These simulations follow standard practice and assume a spatially and temporally constant scaling for the scattering rate as a function of CR rigidity $\nu \sim$ 10$^{-9 }$ (R/GV)$^{-1/2}$ s$^{-1}$ (calibrated explicitly therein to fit all of the observations of CRs in the Milky Way from observations such as Voyager, AMS-02, Fermi, and others).

\subsection{Synchrotron Forward Modeling}\label{sec:synch_methods}
Our methodology to compute the synchrotron specific emissivities from our simulations follows the equations of synchrotron emission as derived in \citet{Ginzburg1965}, and summarized again in \citet{padovani_astronomy_2021}.

For each gas cell in our simulations, we calculate synchrotron emissivities as follows. First, we extract the internally evolved CRe and positron spectra, $j_e(E)$ and components of the magnetic field perpendicular to the line of sight, $\textbf{B}_{\bot}$.

Then, we compute the critical frequency of emission for each spectral bin of CRe,    
        \begin{equation}
        \nu_c(B_{\bot},E) = \frac{3eB_{\bot}}{{4 \pi m_{e}c}} \left(\frac{E}{m_{e}c^{2}}\right)^2
        \end{equation}
where B$_{\rm \bot}$ is the magnetic field strength perpendicular to the line of sight, m$_{\rm e}$ is the electron mass, and c is the speed of light. 

We also calculate the power emitted per unit frequency by CRe in each spectral bin from MeV to TeV energies both parallel and perpendicular to the LOS,

\begin{equation}
        P_{\nu, \bot}(E) = \frac{\sqrt{3}e^{3}}{{2 m_{e}c^{2}}} B_{\bot} [F(x)-G(x)],
\end{equation}

\begin{equation}
        P_{\nu, \|}(E) = \frac{\sqrt{3}e^{3}}{{2 m_{e}c^{2}}} B_{\bot} [F(x)+G(x)],
\end{equation}

where $B_{\bot}$ = $|\textbf{B}_{\bot}|$ and x = $\nu/\nu_c$.

The functions F(x) and G(x) are defined as

\begin{equation}
    F(x) = x \int_{x}^{\infty} K_{5/3}(\xi)d\xi,
\end{equation}

and

\begin{equation}
    G(x) = x K_{2/3}(x),
\end{equation}

where K$_{5/3}$ and K$_{2/3}$ are the modified Bessel functions of order 5/3 and 2/3. For each bin, we use the relevant F(x) and G(x) from pre-computed look-up tables provided by Padovani et al. (priv. comm.). 

We then compute the linearly polarized specific emissivities by integrating the contributions over j$_{e}$,

\begin{equation}
    \epsilon_{\nu,\|} = \int_{m_{e}c^2}^{\infty} \frac{j_{e}(E)}{\mathrm{v_{e}}(E)}P_{\nu, \|}(E) dE,
\end{equation}

and
\begin{equation}
    \epsilon_{\nu,\bot} = \int_{m_{e}c^2}^{\infty} \frac{j_{e}(E)}{\mathrm{v_{e}}(E)}P_{\nu, \bot}(E) dE,
\end{equation}

where $\mathrm{v_{e}}$ is the electron velocity. 

From these specific emissivities, we also compute the Stokes Q$_\nu$ and U$_\nu$ specific emissivities,
\begin{equation}
    \epsilon_{\nu,Q} = [\epsilon_{\nu,\bot} - \epsilon_{\nu,\|}]\cos(2\theta),
\end{equation}
and
\begin{equation}
    \epsilon_{\nu,U} = [\epsilon_{\nu,\bot} - \epsilon_{\nu,\|}] \sin(2 \theta),
\end{equation}
where $\theta$ is the local polarisation angle given by the orientation of $\textbf{B}_{\bot}$ rotated by $\pm$ 90$^\circ$.

When calculating LOS-integrated quantities, we project the galaxy face-on or edge-on using the angular momentum vector of the stars to define the direction perpendicular to the galactic disk, which we define as the $\hat{z}$ direction. All cell vector fields ($\textbf{r}$, $\textbf{B}$, $\textbf{v}$) are transformed accordingly. To compute the corresponding images of the specific intensity I$_\nu$, Q$_\nu$, and U$_{\nu}$, the respective emissivity terms are integrated along the line of sight using a projection routine first described in \citet{Hopkins2005}. This routine appropriately computes the contribution to the emission from every gas cell along the line of sight by taking into account their spatial distribution and hydrodynamic smoothing lengths.

\section{RESULTS: FIRE Simulations with Resolved Cosmic Ray Spectra}\label{sec:results}

\subsection{Synthetic Observations}
\begin{figure*}
    \centering
    \includegraphics[width=1.0\textwidth]{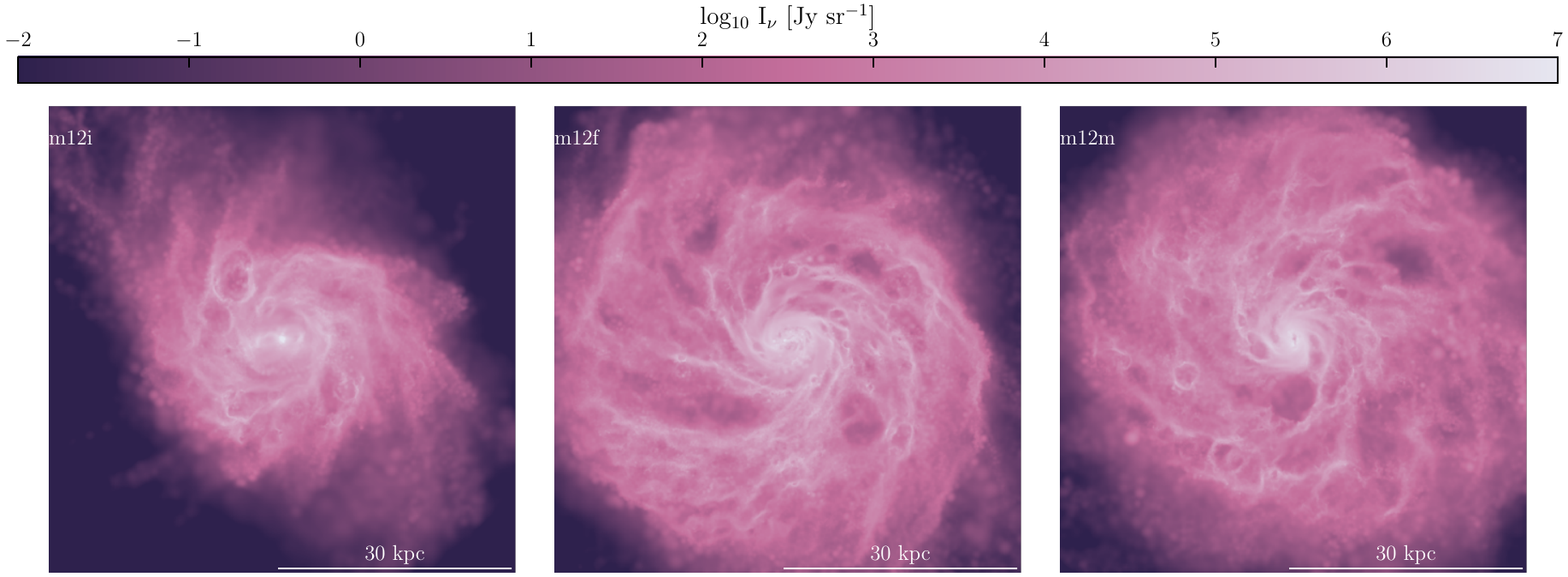}
    \caption{\textit{Images of specific intensity (I$_\nu$)} for \texttt{m12i}, \texttt{m12f}, and \texttt{m12m} at $\lambda$ = 6.2 cm, with the colorbar showing log$_{10}$(I$_\nu$). All three simulations exhibit a range of I$_\nu$ values broadly consistent with radio continuum observations of nearby spiral galaxies. The synchrotron intensity closely traces the neutral gas spatial distribution, with enhanced emission coincident with spiral structure and relatively lower levels of emission in inter-arm regions.}
    \label{fig:m12s_viz}
\end{figure*}

In this section, we present our synthetic observations and explore which ISM phases contribute to the resultant emission. 
We first show images of I$_\nu$ at $\lambda$ = 6.2 cm for face-on projections of the galactic disks for the three fiducial simulations, following the procedure outlined in Section \ref{sec:synch_methods}, in Figure \ref{fig:m12s_viz}. The images are in their fiducial high-resolution format at a scale of $\sim$120 pc/pix and not smoothed to an observational beam. 

A few things are evident from visual inspection: first, our synthetic synchrotron images have typical values of I$_\nu$ in qualitative agreement with those observed in nearby spiral galaxies \citep{basu_magnetic_2013,beck_magnetic_2015} with spatially resolved radio continuum observations; we defer quantitative comparisons to the observations to Section \ref{sec:obs_comp}. Second, the synchrotron emission is broadly coincident with the spatial distribution of neutral gas, with stronger emission coming from dense gas near the galactic center and correlated with spiral structure, and weaker emission in inter-arm regions and towards the galactic outskirts.

\begin{figure}
    \centering
    \includegraphics[width=0.5\textwidth]{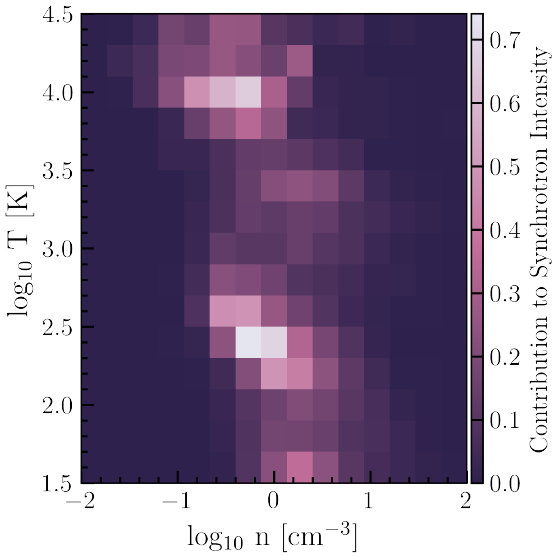}
    \caption{\textit{ISM phase diagram, or 2-D histogram of temperature vs. density of gas cells}, weighted by contribution to specific synchrotron intensity for \texttt{m12f} at $\lambda$ = 6.2 cm, within cylindrical R $<$ 10 kpc. Color-bar shows the probability density. Consistent with visual inference from Figure \ref{fig:m12s_viz}, the synchrotron emission is largely dominated by neutral gas typical in CNM and WNM ISM conditions, rather than diffuse, volume-filling, and ionized phases.}
    \label{fig:m12f_hist}
    
\end{figure}

In Figure \ref{fig:m12f_hist}, we present 2-D histograms of the gas density and temperature, for the gas cells producing the emission visualized in Figure \ref{fig:m12s_viz}, all weighted by each gas cell's contribution to the specific synchrotron intensity \Inu. In these intensity-weighted ISM phase diagrams, the bimodal nature of gas which dominates the emission becomes clear; the synchrotron emission primarily arises from the cold, neutral medium (CNM - T $\sim$100-300 K, n $\sim$3-30 cm$^{-3}$) and warm, neutral medium (WNM - T $\sim$3$\times \rm{10^{3}}$ -10$^{4}$ K, n $\sim$0.1-1 cm$^{-3}$). Inspecting similar intensity-weighted histograms of the gas cells' neutral hydrogen fractions confirms that this gas is primarily neutral.

This evinces a physical scenario in which relatively cool, dense gas, which comprises a small fraction of the total ISM volume, contributes significantly to synchrotron emission. While warm, more appreciably volume-filling ISM gas is synchrotron-bright as well, the emission arises largely from the denser and neutral warm gas as opposed to the more diffuse (n $<$ 0.1 cm$^{-3}$), ionized warm/hot gas, which fills the majority of the ISM volume. Put more quantitatively, neutral gas with (n $>$ 0.1 cm$^{-3}$, T $<$ 10$^{4}$ K) contributes $\sim$80\% of the emission, despite filling only $\sim$20\% of the volume. 

We note that a scenario in which CRs are not modeled self-consistently would predict a different intensity-weighted phase distribution: if one assumed a priori that e$_{\rm CR}$ was locally in equipartition with a self-consistently evolved B, then the intensity-weighted distribution would be more biased towards higher density gas as \textbf{B} $\sim$ $\rho^{\alpha}$, with $\alpha \sim$ 0.5-0.66 \citep{Ponnada2022}. Whereas if one assumed the ISM to be a homogeneous slab with a constant volumetric B in equipartition with CRs, the distribution would then effectively look like the volume-weighted distribution and thus weighted towards the more diffuse phases of the ISM. This result subsequently has physical implications for what traditional equipartition models \`{a} la \citet{beck_revised_2005,lacki_equipartition_2013} infer in the form of the "mean \textbf{B}" from synchrotron observations, which we show and discuss in Sections \ref{sec:equi}, \ref{sec:toy_model}, and \ref{sec:discussion}. 

\subsection{Radial intensity profiles and comparisons to observations}\label{sec:obs_comp}

\begin{figure}
    \centering
    \includegraphics[width=0.5\textwidth]{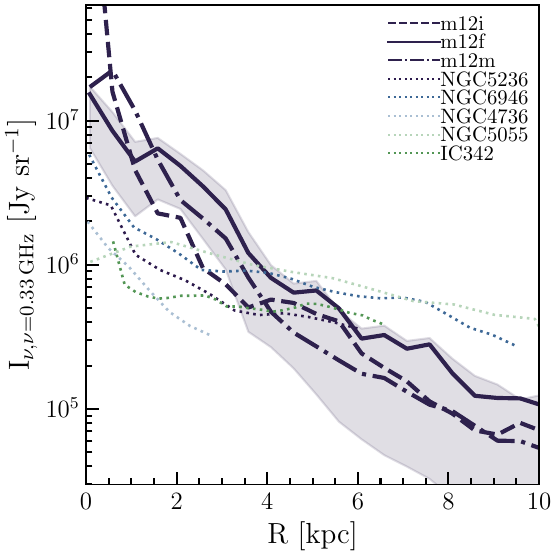}
    \caption{\textit{Azimuthally averaged (mean) radial profiles of synchrotron specific intensity at 0.33 GHz} for \texttt{m12i}, \texttt{m12f}, and \texttt{m12m}, in navy dashed, solid, and dot-dashed lines. Shaded regions show the 25-75 percentile range at a given radial bin. Corresponding radial profiles of non-thermal synchrotron emission (free-free emission corrected) for nearby face-on spiral galaxies from \citet{basu_magnetic_2013} and \citet{beck_magnetic_2015} are shown in dotted lines, with 4.83 GHz observations of IC342 scaled up by $\nu/\nu_{0}$$^{\alpha}$. Our synthetic synchrotron images for the fiducial model agree to within an order-of-magnitude with spatially resolved synchrotron emission from nearby spiral galaxies. Variation in the shapes of the profiles towards the galactic centers and outskirts partially arises due to differences in the gas surface density (see Figure \ref{fig:norm_profiles} for corresponding plot), though the simulations appear to produce steeper radial profiles in comparison to the observed systems compared.} 
    \label{fig:rad_profiles}
\end{figure}

In Figure \ref{fig:rad_profiles}, we compare the radial profiles of synchrotron specific intensity at observing frequencies of 0.33 GHz from our fiducial simulations to a few observed face-on spiral galaxies of roughly similar mass. We have also calculated the simulations' emission profiles at 4.8 GHz, as one of the observed galaxies, IC342, was measured at this frequency. But since the relative comparison is similar, we simply plot all systems at 0.33 GHz, shifting IC342 according to the authors' assumed spatially constant spectral index of -1. The radial profile of the non-thermal emission as plotted in \citet{beck_magnetic_2015} for IC342 is also shown, and for the other observed galaxies, we show the radial profiles of the non-thermal emission (corrected to remove the free-free emission) out to the radii published in \citet{basu_magnetic_2013}. 

We find the predicted \Inuhi for our simulations generally ranging from 10$^{5.5}$ to $\sim$10$^{4}$ Jy/sr at the galactic center to the "Solar circle" (galacto-centric cylindrical radius R = 7-9 kpc) and \Inulow ranging from 10$^{7}$ to $\sim$10$^{5}$ Jy/sr in Fig. \ref{fig:rad_profiles}.  When comparing our predicted radial emission profiles to observations of nearby spiral galaxies, we see an order-of-magnitude agreement, though two of the four observed profiles (NGC5055 and NGC 6946) appear to fall more slowly at large R $\geq$ 5 kpc. We caution that none of these simulations are meant to be exact analogs of the observed galaxies, but are only order-of-magnitude similar in morphology and galaxy mass. Various galaxy-to-galaxy differences in the profiles partly reflect this set of galaxies' diversity in properties like gas surface density and star formation rate, and in Figure \ref{fig:norm_profiles} we show similar specific intensity profiles normalized to the gas surface density, which are qualitatively more similar in the shape of the profiles.

\subsection{Spectral variation and emission properties}
Evolving the full CR spectrum is computationally expensive, and most simulations of galaxy formation including CRs utilize the `single-bin' approximation, solely evolving the 1-10 GeV e$_{\rm CRp}$. We are therefore motivated to test whether explicitly evolving CR spectra makes significant quantitative or qualitative differences in the predicted synchrotron emission. 

\begin{figure}
    \centering
    \includegraphics[width=0.45\textwidth]{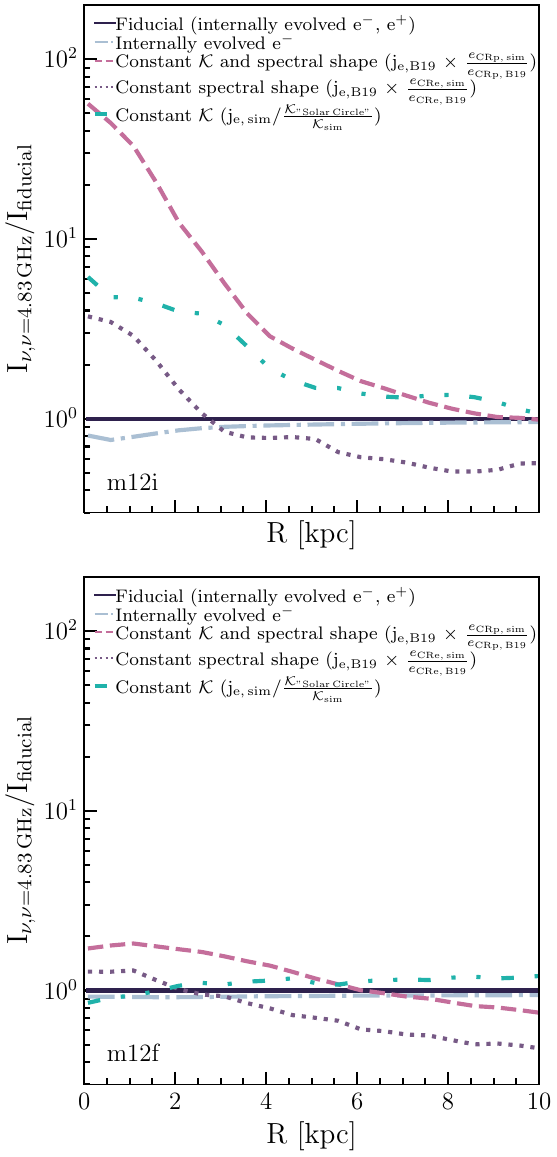}
    \caption{\textit{Azimuthally averaged (mean) radial profiles of the ratio of I$_{\nu}$ to I$_{\nu,\, fiducial}$ under different assumptions for j$_{\rm e}$} for \texttt{m12i} (top) and \texttt{m12f} (bottom) at 4.83 GHz. Lines show fiducial intensity using internally evolved j$_{\rm e}$ including e$^{+}$ contribution (navy solid), fiducial intensity without e$^{+}$ (light blue dot-dashed), holding spectral shape and $\mathcal{K}$ constant by re-scaling j$_{\rm e}$ of \citet{bisschoff_new_2019} (B19) to internally evolved e$_{\rm CRp}$ (pink dashed), holding the spectral shape constant by re-scaling the fiducial spectrum by e$_{\rm CRe}$ (purple dotted), and holding $\mathcal{K}$ constant by re-scaling the fiducial j$_{e}$ to the nominal value at that simulation's "Solar Circle" (teal). Emission from secondaries contributes fractionally to overall emission at all radii, though reaching $\sim$20\% contribution at high $\Sigma_{\rm gas}$. As demonstrated in \texttt{m12i}, strong CRe cooling at relevant energies corresponds to over-predicting emission towards the galactic center, where $\Sigma_{\rm gas}$ is relatively higher, for constant $\mathcal{K}$ and/or spectral shape assumptions. This effect driven by variation in $\mathcal{K}$ compounding with spectral shape effects as shown by the pink line. Thus, in galactic environments where losses become important, modeling the CR spectra and $\mathcal{K}$ explicitly is necessary for predictive calculations, while in environments where cooling is relatively weak at relevant energies (galactic outskirts and much of \texttt{m12f}), spectral shape/$\mathcal{K}$ assumptions make at most factor of $\sim$2 differences.}

    \label{fig:spec_var}
\end{figure}

To compare what the emission would look like for the same simulated galaxies if we did not evolve the full spectral information of CRs, we illustrate a few different spectral assumptions in Figure \ref{fig:spec_var} starting from our fiducial simulations of \texttt{m12i} and  \texttt{m12f}. First, we see that excluding emission from positrons (light-blue, dot-dashed lines) deducts minimally, contributing $\lesssim$ 10\% of the emission at most radii, and reaching $\sim$20\% near the galactic center of \texttt{m12i}.

Secondly, we show the results corresponding to scaling the CR electron spectrum of \citet{bisschoff_new_2019} by e$_{\rm CRe,\, sim}$, thereby holding the shape of j$_{\rm e}$ constant (purple dotted lines), as well as scaling according to e$_{\rm CRp}$ in each gas cell, which additionally fixes $\mathcal{K}$ (pink dashed lines), the total energy density ratio e$_{\rm CRp}$/e$_{\rm CRe}$ to the literature value (akin to what might be done in the "single-bin" scenario where only e$_{\rm CRp}$ is known). Holding the spectral shape constant with the "right" $\mathcal{K}$ shows that one would tend to over-predict the emission by up to factors of $\sim$6 in regions where losses at the energies of interest are strong (towards the galactic center). Additionally fixing $\mathcal{K}$, we see that evolving the CR spectral information can make a modest difference at a factor of $\sim$2 in \Inu for typical spiral galaxy conditions as shown in much of $\texttt{m12f}$ and the outer radii of $\texttt{m12i}$, but can make large (factor of $\sim$10-50) differences towards the central regions of the galaxy. 

We break this down in more detail, showing that these deviations owe to a radially-varying proton-to-electron ratio between the galactic center and outskirts compounding with changing spectral shape (comparing constant spectral shape and constant $\mathcal{K}$ + spectral shape lines, which differ only in normalization) to further over-predict the emission, rather than the varying proton-to-electron ratio alone (green dashed-dotted lines, constant $\mathcal{K}$, j$_{\rm e, \, sim}$). This effect arises in \texttt{m12i} due to a quasi-starburst in the circum-nuclear region R $<$ 3 kpc, where $\Sigma_{\rm gas}$ quickly rises from $\sim$30-250 $M_{\odot}$ pc$^{-2}$ in comparison to lower surface densities $\sim$30-70 and 30-100 $M_{\odot}$ pc$^{-2}$ in \texttt{m12f} and \texttt{m12m} respectively. Correspondingly, loss terms owing to higher gas densities being correlated with higher \textbf{B}, increased radiation field intensities, etc., result in significant cooling of the CRe spectra at the energies of interest and a changing $\mathcal{K}$.  

As expected, there is little difference at "Solar circle" radii, where our simulations have been shown to faithfully agree with the observed LISM spectrum \citep{hopkins_first_2022}. However, these results illustrate the need to self-consistently model the CR spectral shapes and $\mathcal{K}$ to accurately predict synchrotron emission in galactic environments where loss terms become important, as simple spectral assumptions from a 'single-bin' treatment of CR protons can lead to order-of-magnitude differences in the predicted emission. These differences may also vary with frequencies different from the $\sim$GHz observations probed here which would probe different energy intervals of the CRe spectrum. Furthermore, evolution in $\mathcal{K}$ has implications on key assumptions invoked in equipartition estimates of \textbf{B}, as we detail in the next section.

\subsection{Equipartition magnetic field strengths: what do they really measure?}\label{sec:equi}
The most commonly used formalism to determine equipartition estimates of the magnetic field (\textbf{B$_{\rm eq}$}) from radio continuum observations of non-thermal emission from galaxies is that of \citet{beck_revised_2005}. The equation is as follows:

\begin{multline}\label{eq:equi}
    \textbf{B}_{\rm{eq}} = \{ 4\pi(2\alpha+1) (\textbf{K}_{0}+1) I_{\nu}E_{p}^{1-2\alpha}(\nu/2c_{1})^{\alpha} \\ /\left(2\alpha-1)c_{2}(\alpha)(f_{V}*L)c_{4}(i)\right\} ^{1/(\alpha+3))} \}
\end{multline}

where $\alpha$ is the synchrotron spectral index, E$_{\rm p}$ is the proton rest energy, and c$_{1}$, c$_{2}$, and c$_{4}$ are combinations of physical constants which encapsulate dependencies on $\alpha$ and the inclination of the magnetic field. The equipartition formula also requires assumptions about the depth of the emitting material, L (here the implicit dependence of the volume-filling factor of emitting gas, f$_V$, is written explicitly), and \textbf{K$_{0}$}, the ratio of number densities of CRp and CRe at energies from E$_{\rm p}$ up to some energy E$_{\rm syn}$ beyond which synchrotron and IC losses dominate for CR electrons\footnote{Note that \ko differs from $\mathcal{K}$, which is the ratio of the total CRp/CRe energy densities.} \citep{beck_revised_2005}. 

As described in the Introduction, equipartition estimation of the magnetic field is subject to a few extra assumptions beyond that of e$_{\rm CR}$ = u$_{\rm B}$. The first auxiliary assumption is that of a constant ratio of the number densities of CRe and CRp, \ko $\sim$100, motivated by the injection spectrum of primary CRs from SNe via diffusive shock acceleration and measurements at the Milky Way Solar Circle \citep{Bell1978,Bell2004,beck_revised_2005}. Though there have been modifications to this assumption for galaxies where the synchrotron emission is expected to have significant contribution from secondary CRe and positrons generated from the resultant pion-decays of primary CR collisional losses \citep{lacki_equipartition_2013}, these are not generally applied to galaxies like those in Fig. \ref{fig:rad_profiles}. This assumption further implicitly requires that the CRp and CRe spectrum to have constant and equal power-law indices, which holds close to injection sites, but may not hold in a spatially and temporally independent manner across galaxies. It also assumes negligible contribution from positrons, and that the synchrotron is optically thin.

Secondly, CRe and positrons, which dominate the synchrotron emission at frequencies of a few GHz, and $\sim$1-10 GeV CR protons, which dominate overall the energy density, are assumed to have the same, spatially and temporally uniform distribution within and across each pixel, and as a function of height above the mid-plane up to some height L. These assumptions are subject to questioning on galactic scales, where the dominant energy loss terms and correspondingly the loss timescales for CRe and CR protons can differ significantly due to large local variations (along a line of sight and across an observational beam) in gas density, magnetic and radiation energy densities, and phase structure \citep{wolfire_neutral_1995,Evans1999}. 

Thirdly, the size of the emitting region both along the line of sight and within a given pixel or beam must be assumed. While traditional applications of the equipartition formula to galaxies have assumed path lengths of L $\sim$1-2 kpc for face-on observations, it remains unknown what the typical volume filling factor of synchrotron emitting gas actually is in galaxies. Implicit within this assumption is a homogeneous and volume-filling field, which given the multi-phase nature of the ISM, may not reflect the true \textbf{B} that primarily contributes to the synchrotron emission.

From the synthetic synchrotron images and radial profiles shown in the previous sections, we can now explore equipartition magnetic field strengths (\textbf{B$_{\rm eq}$}) from the forward-modeled specific synchrotron intensity and test the several assumptions invoked. 

\begin{figure*}
    \centering
    \includegraphics[width=1.0\textwidth]{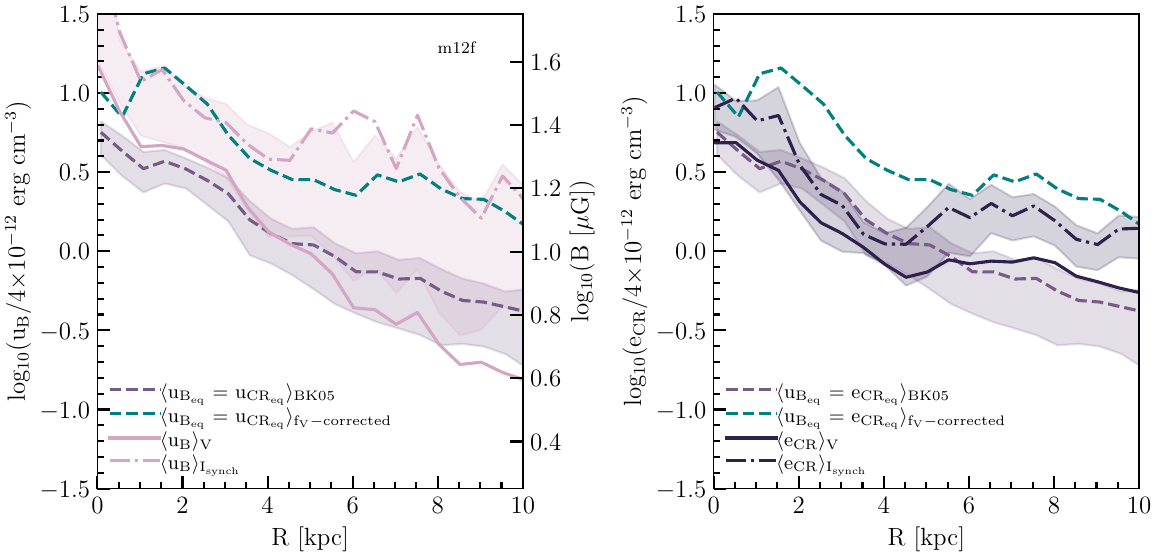}
    \caption{\textit{Azimuthally averaged radial profiles of u$_{\rm B}$ (left) and e$_{\rm CR}$ (right) weighted by volume and I$_{\nu}$ in comparison to u$_{\rm B_{eq}}$} for \texttt{m12f}. Lines represent u$_{\rm B}$ (pink), e$_{\rm CR}$ (navy), volume-weighted (solid) and \Inulow-weighted (dot-dashed)  averages of each quantity, within $|$z$|$ $\leq$ 0.5 kpc. The result of applying Equation \ref{eq:equi} to the radial profile presented in Figure \ref{fig:rad_profiles} is shown, with L = 1, \textbf{K}$_{\rm 0}$ = 100, f$_{\rm V}$ = 1 (purple dashed) and f$_{\rm V}$ computed within each radial bin (teal dashed). Shaded regions show the approximate $\pm$ 1$\sigma$ scatter (32-68 percentile) of the emission-weighted quantity and fiducial equipartition model at a given radial bin. Unilaterally, we see that without correcting for the volume-filling factor of the neutral mid-plane gas which dominates \Inu, \textbf{the equipartition formula under-predicts the "true" \Inu-weighted \textbf{B} by $\sim$0.3-0.6 dex and e$_{\rm CR}$ by  by $\sim$0.1-0.6 dex}. Furthermore, due to the emission being dominated by mid-plane gas, the equipartition model with the right volume-filling factor is generally not representative of the volume-weighted u$_{\rm B}$, though the fiducial model surprisingly traces the volume-weighted quantities well owing to a confluence of factors (see Section \ref{sec:toy_model}). While not shown in this figure, \texttt{m12i} and \texttt{m12m} show generally the same behavior.}. 
    \label{fig:equipartition_comparison}
\end{figure*}

In Figure \ref{fig:equipartition_comparison}, we show  estimators of \textbf{B} weighted by \Inulow and volume in \texttt{m12f}, and how \Beq compares. We show \Beq resulting from assuming $\alpha$ = 1, L = 1 kpc, \ko = 100, and f$_{V}$ = 1, which are typical model assumptions in observational literature. In this case, it becomes immediately clear that Equation \ref{eq:equi} \textit{under-predicts the "true" \Inu-weighted \textbf{B} by $\sim$0.3-0.6 dex across all radii and e$_{\rm CR}$ by $\sim$0.3-1.0 dex for R $>$ 4 kpc} (though interestingly, traces the volume-weighted \textbf{B} well, which we discuss in detail in Section \ref{sec:toy_model}). Examining the assumptions for \ko and f$_{V}$ in this model, we find \ko varies from $\sim$100 $\pm$ 10 near the galactic center to $\sim$60 $\pm$ 10 near the "Solar circle" with the exception of \texttt{m12i}, where \ko rises rapidly to $\sim$10$^{3}$ near the galactic center due to strong leptonic losses at relatively higher gas surface densities $\gtrsim$ 150 M$_{\odot}$ pc$^{-2}$. Varying the assumed value of \ko according to this radial variation would only affect the equipartition-inferred values at the tens of percent level (\Beq $\sim$\ko$^{1/4}$), and so we find that \ko = 100 is a decent assumption in our simulations for most of the galaxy conditions sampled here, with the exception of the very inner (R $<$ 2 kpc) region of \texttt{m12i}. 

The assumption of the depth of the emitting material being $\sim$1 kpc is more suspect, however. Computing the volume filling fraction of the gas cells which contribute to the upper 50$\%$ of \Inu within radial bins of vertical thickness 1 kpc reveals that f$_{V}$ varies from $\sim$0.05 - 0.2 over the radial range. This volume filling fraction is very similar to the ratio of the face-on scale height of emission to the path length, H$_{I_{\nu}}$/L, implying that the emission is primarily arising from the mid-plane. This, in conjunction with Figure \ref{fig:m12f_hist}, further develops the picture of the emission primarily being dominated by the cold and warm neutral medium relatively confined to the thin disk.

When f$_{V}$ is subsequently corrected to representative values $\sim$0.05 - 0.2, the equipartition formula gives values of \Beq and e$_{\rm CR}$ that are closer to the "true," \Inu-weighted values of \textbf{B} (though over-predicting \Inu-weighted e$_{\rm CR}$ owing to deviations from physical equipartition with u$_{\rm B}$). The \Inu-weighted \textbf{B} is generally higher (0.2-0.6 dex) than the volume-weighted \textbf{B}, while the \Inu-weighted e$_{\rm CR}$ and volume-weighted  e$_{\rm CR}$ exhibit less difference. This indicates that the primary effect of correcting f$_{V}$ is to correct for the higher \textbf{B} in the denser mid-plane gas, rather than to correct for e$_{\rm CR}$, which more weakly scales with gas density in our simulations owing to CRs' ability to diffuse \citep{Ponnada2022,hopkins_first_2022}. It is important to note that the volume filling gas at heights L $\sim$ 1 kpc above the mid-plane at R $\sim$ 5 kpc has typical \textbf{B} $\sim$2-4 $\mu$G $\ll$ $\langle \textbf{B}_{\rm synch} \rangle$, \citep[see][for details]{Ponnada2022}, so the traditional estimator also severely over-estimates typical \textbf{B} in more diffuse gas above the disk, via assumptions of a homogeneous, height-independent \textbf{B}.

 We emphasize that this is consistent with state-of-the-art, high-resolution radio continuum observations \citep{krause_chang-es_2018, heesen_radio_2018}; detailed studies of nearby edge-on spiral galaxies have revealed very bright thin-disk components with scale heights ranging from 10s-100s of pc, which are $\lesssim$ the observational beam size and thus subject to large errors. The best-fit two-component models to these observations all indicate that this thin, mid-plane component almost ubiquitously dominates the overall emission, with additional extended thick-disk components with nominal scales of L $\sim$kpc as invoked in equipartition models contributing fractionally to the overall emission. Thus, when viewed face-on, the emission largely traces the thin disk component, and so taking typical values for L associated with these extended synchrotron halos rather than the scale height of the medium which contributes most to the emission (which is still subject to a high degree of observational uncertainty), leads to vastly under-predicting the \Inu-weighted \textbf{B}. 
 
 We have checked that treating edge-on synchrotron images of our simulations in an observational manner, smoothing to a representative observational beam of 10", and fitting two-component exponential disk models akin to observational studies yield similar results for thin and thick disk scale heights ($\sim$100 pc, $\sim$1 kpc respectively; see Figure \ref{fig:scale_heights}).

\section{A Toy model for estimating B from synchrotron emission in a multi-phase medium}\label{sec:toy_model}
In this section, we present a toy model to characterize \textbf{B} in a multi-phase medium to gain intuition as to what factors cause the traditional model to deviate from the "true" values, given the understanding from the previous section that the synchrotron emission is primarily arising from neutral mid-plane gas (up to variations in the CR transport physics).

Consider the mean, volume-weighted magnetic energy density in a vertical ISM slab (at a given cylindrical radius R) of some finite height to be described by an exponential function anchored to the mean mid-plane ($|z| \leq$ 0.1 kpc) value of u$_{\rm B}$:

\begin{equation}\label{eq:uB_profile}
    \langle u_{\rm B} \rangle = \langle u_{\rm B_{0}} \rangle e^{-(|z|/H_{B})} 
\end{equation}

where the radial dependencies of $\langle u_{\rm B_{0}} \rangle$ and H$_{\rm B}$ (the magnetic scale height) are implicit. Correspondingly, we can describe the volume-weighted average of the CR energy density in each slab:
\begin{equation}\label{eq:uCR_profile}
     e_{\rm CR}  = \psi * u_{\rm B_{0}} \Bigr(\frac{\langle u_{\rm B} \rangle}{\langle u_{\rm B_{0}} \rangle}\Bigr)^{\beta}
\end{equation}

where $\psi$ is $ \langle e_{\rm CR_{0}} \rangle$/$\langle u_{\rm B_{0}} \rangle$ and accounts for differences in the energy density at the mid-plane\footnote{For a more detailed look at $\psi$, which describes the volume-averaged physical "equipartition" between \textbf{B} and CRs, we present radial profiles of $\psi$ in Figure \ref{fig:psi_profile}.} and $\beta$ accounts for differences in the vertical scale height of u$_{\rm B}$ and e$_{\rm CR}$ and the local inter-dependence. 

Within each vertical slab, we can make the assumption that the coherence length of the magnetic field is much less than the vertical scale height or the radius, and thus construct a volume-weighted PDF of given magnetic fluctuations ($\delta_{u_{B}}$ = $u_{\rm B}/ u_{\rm B_{0}}$) describing the probability of a given unit volume having a magnetic fluctuation $\delta_{u_{B}}$. We further make the ansatz that this PDF is a log-normal distribution of fluctuations in $\delta_{u_{B}}$, motivated by descriptions of density fluctuations in MHD turbulence \citep{hopkins_model_2013,beattie_density_2022}:

\begin{equation}\label{eq:pdf}
P_{V}(ln \, \delta_{u_{B}}) = \frac{V_{tot}}{\sqrt{2 \pi S_{\delta}}} \exp \Bigl\{-   \frac{(ln \, \delta_{u_{B}} + S_{\delta}/2)^{2}  }{2S_{\delta}} \Bigr\}
\end{equation}
 where V$_{\rm tot}$ is the total volume within a vertical slab, S$_{\delta}$ describes the volume-weighted variance of ln($\delta_{u_{B}}$) arising from turbulent, multi-phase structure.

 With these equations in hand, we can build a model which relates the observable synchrotron intensity (averaged over some finite pixel resolution A) to the underlying volume-weighted \textbf{B}. 
 
 Starting with the same assumptions as in \citet{beck_revised_2005} and including spectral dependencies implicitly ($\alpha$ will remain a free parameter), one can work to an expression for $\langle$\Inu$\rangle$ given Equations \ref{eq:uB_profile} and \ref{eq:uCR_profile} as follows:

 \begin{equation}
     \langle I_{\nu} \rangle =  \frac{\int\int \psi C(\alpha) u_{B}^{\gamma} dA dz}{A} = \frac{1}{A}\int \psi C(\alpha) u_{B}^{\gamma}  u_{\rm B_{0}}^{1- \beta} dV
 \end{equation}
 
 where dV is a volume element, C($\alpha$) is a combination of physical constants with minor dependencies on the power-law index and $\nu$\footnote{    $C_{\alpha} = \frac{(2 \sqrt{2\pi})^{\alpha+1} c_{4}^{2} (4\alpha-2)c_{2}(\alpha)}{(\textbf{K}_{0}+1)(2\alpha +1)(\frac{\nu}{2c_{1}})^{\alpha} E_{p}^{1-2\alpha}   }$, where c$_{1}$-c$_{4}$ are defined in Appendix A of \citet{beck_revised_2005}}, and $\gamma$ = [$\frac{(\alpha+1)}{2}$ + $\beta$]. From this expression, we can utilize the PDF given by Equation \ref{eq:pdf} to find a closed form expression for $\langle I_{\nu} \rangle$
as follows:

\begin{equation}\label{eq:toy_Inu}
    \langle I_{\nu} \rangle = \frac{2\psi C * H_{B} \exp \Bigl\{\frac{S_{\delta}(\gamma -1)\gamma}{2}\Bigr\}}{\gamma}  \langle u_{\rm B_{0}}\rangle^{\frac{(\alpha+3)}{2}}
\end{equation} 

which can be re-arranged to find an expression for $\langle u_{\rm B_{0}} \rangle$:

\begin{equation}\label{eq:toy_uB}
    \langle  u_{\rm B_{0}} \rangle = \left(\frac{\gamma \langle I_{\nu} \rangle}{2\psi C * H_{B} \exp \Bigl\{\frac{S_{\delta}(\gamma -1)\gamma}{2}\Bigr\}}\right) ^{2/(\alpha+3)}
\end{equation}
 
 This expression is analogous to the  \citet{beck_revised_2005} formalism of Equation \ref{eq:equi}, but relaxes the assumptions of equipartition between u$_{\rm B}$ and e$_{\rm CR}$ and a homogeneous medium with uniform \textbf{B} and e$_{\rm CR}$; instead, we parameterize our ignorance of equipartition into a simple power-law relation between the two quantities and variations due to inhomogeneity in the ISM in the form of the volume-weighted variance in $\delta_{u_{B}}$ and the magnetic scale height H$_{\rm B}$. Our revised expression reduces to Equation \ref{eq:equi} when S$_{\delta}$ = 0, $\psi$ = 1, and $\beta$ = 1.

Note that our Eq.~\ref{eq:toy_uB} for $\langle u_{\rm B_{0}} \rangle = \langle u_{\rm B} \rangle_{V}$ is identical to the \citet{beck_revised_2005} (BK05) formula in Eq.~\ref{eq:equi} with the replacement $f_{\rm V} \rightarrow \psi\,f_{\rm cl}\,H_{\rm B}/L$ (BK05 implicitly take $f_{\rm V}=1$), where $f_{\rm cl} \equiv \gamma^{-1}\,\exp{[\gamma\,(\gamma-1)\,S_{\delta}/2]}$.

We have three terms here parameterizing three physical assumptions/uncertainties: (1) $\psi$ represents the (mean) deviation from equipartition; (2) $f_{\rm cl}$ represents the "clumping factor" which can boost emission (for a given volume-weighted set of properties) owing to substructure with larger magnetic and/or CR energy density; and (3) $H_{B}/L$ simply corrects the ad-hoc constant $L=1\,{\rm kpc}$ to the "correct" size of the emitting disk. Observationally, of course, these are largely unknown, hence taking $f_{V}=1$ in BK05. But here, we can estimate the true values in the simulations of each parameter. 

The value of $\psi$ can be directly read off from Fig.~\ref{fig:equipartition_comparison}, increasing from $\sim 0.7-3$ at $R\sim1-10\,$kpc. This radial trend is expected given the diffusive nature of CRs (see discussion of this in \citealt{hopkins_first_2022}), since it means the CR energy density will fall less-rapidly than the gas pressure (and $u_{\rm B}$) at large galacto-centric radii. Taking $H_{\rm B}$ to be the atomic disk scale-height gives $H_{\rm B} \sim 200\,$pc, similar to canonical scale heights for the star-forming disks of the Milky Way and other observed galaxies \citep{Kalberla2009,Yim2014,Patra2020,Gensior2023}. 

Direct examination of the $e_{\rm CR}-u_{\rm B}$ correlations (shown in \citealt{hopkins_first_2022}) gives $\beta \sim 0.2-0.3$: this arises again because the CRs are diffusive, so (especially on small scales) always form a locally-smooth distribution compared to the magnetic fields \citep{Butsky2018,chan_cosmic_2019,buck_effects_2020}. And we can estimate $S_{\delta}$ directly from the midplane scatter in $\ln{u_{\rm B}}$ giving $S_{\delta}\sim 2-4$ -- this is naturally expected from the same turbulence models which motivated our lognormal assumption in the first place, which predict $S_{\delta} \approx \ln[1 + \mathcal{M}_{c}^{2}]$ (where $\mathcal{M}_{c}$ is the compressive Mach number of ISM turbulence, and $\mathcal{M}_{c} \sim $\,a few in both observations and simulations; see \citealt{Federrath2010,hopkins_model_2013}).

 We can immediately follow the same procedure to calculate the intensity-weighted magnetic energy density, 
 \begin{equation*}
     \langle u_{\rm B}\rangle_{I_{\nu}} = (\gamma/(\gamma+1))\,\exp{(\gamma\,S_{\delta})}\,\langle u_{B_{0}}\rangle,
 \end{equation*}

as well as the volume-weighted and intensity-weighted midplane CR energy densities 

\begin{equation*}
    \langle e_{\rm CR} \rangle_{V} = \psi\,\langle u_{\rm B_{0}} \rangle,  
\end{equation*}

\begin{equation*}
        \langle e_{\rm CR}\rangle_{I_{\nu}} = (\gamma/(\gamma+\beta))\,\exp{[\beta\,(\alpha+3\,\beta)\,S_{\delta}/2]}\,\langle e_{\rm CR}\rangle_{\rm V}.
\end{equation*}

This provides a natural explanation for several phenomena we saw in Fig.~\ref{fig:rad_profiles}. The ratio $\langle u_{\rm B}\rangle_{I_{\nu}} / \langle u_{\rm B}\rangle_{\rm V} \sim 30-40$ in the outer disk, arises primarily from the "clumping factor" correction. The ratio of $\langle u_{\rm B} \rangle_{\rm V} = \langle u_{\rm B_{\rm 0}}\rangle$ in the disk mid-plane to $\langle u_{\rm B} \rangle_{\rm BK05}$ (the standard equipartition formula) is given by $(\psi\,f_{\rm cl}\,H_{B}/L)^{-2/(\alpha+3)}$, which in the outer disk is $\sim 0.7$ (and is $\sim 1.4$ in the inner disk) so the BK05 formula will slightly over-estimate (under-estimate) $\langle u_{\rm B} \rangle_{\rm V}$ in the outer (inner) disk, as we see. The difference is small -- i.e.\ BK05 appears to "correctly" obtain roughly the correct $\langle u_{\rm B} \rangle_{\rm V}$, because the corrections $f_{\rm cl}$ and $H_{\rm B}/L$ are both relatively large but tend to go in opposite directions (cancelling each other out), combined of course with the effect of the small power law $2/(\alpha+3)$ which tends to suppress any differences. In other words, the true emitting region is smaller along the line of sight than assumed by BK05, but the emission is also boosted within that region by clumping. Of course, BK05 under-estimates $\langle u_{\rm B}\rangle_{I_{\nu}}$ by a very large factor as predicted. 

Using the same exercise/model we can compare the intensity and volume-weighted CR energy densities $\langle e_{\rm CR}\rangle_{I_{\nu}}$ and $\langle e_{\rm CR}\rangle_{\rm V}$. Given the small $\beta \ll 1$, we predict that although $\langle u_{\rm B}\rangle_{I_{\nu}} \gg \langle u_{B}\rangle_{\rm V}$, $\langle e_{\rm CR}\rangle_{I_{\nu}}$ only exceeds $\langle u_{\rm CR}\rangle_{\rm V}$ by a factor $\sim 1.5-2$. Thus for a given $\psi$, $\beta$ and $S_{\delta}$ or $f_{\rm cl}$, and $H_{\rm B}$, we can quantitatively explain all of the relative values of the different estimators in Fig.~\ref{fig:rad_profiles}.

\section{Summary and Conclusions}\label{sec:discussion}
In this work, we have presented the first end-to-end predictions of synchrotron emission from MHD galaxy formation simulations which self-consistently evolve \textbf{B} and the CR(e) spectra from \citet{hopkins_first_2022}. These simulations utilize a constant in space and time scaling for the CR scattering rate $\nu$, calibrated to reproduce both Solar System CR data (e.g. Voyager, AMS-02) and resolved $\gamma$-ray observations of the MW and nearby galaxies (e.g. Fermi).

 We have found that synchrotron emission in L$_{*}$ galaxies arises not only from the volume-filling, warm/hot phases of the ISM, but can be dominated by cooler and denser WNM/CNM phases. This is not unexpected; the long discovered and well studied FIR-Radio correlation of galaxies \citep{voelk_radio_1985,ivison_radio_1985,de_jong_radio_1985,helou_thermal_1985,condon_correlations_1991} which exists also on sub-kpc resolved scales \citep{murphy_initial_2006} requires a connection between the FIR emission, which arises from dust re-radiating UV photons from star formation, and the synchrotron emission which also arises from star forming regions with high neutral gas densities and magnetic field strengths. Furthermore, recent synchrotron observations of edge-on galaxies \citep{krause_chang-es_2018,heesen_radio_2018} have found bright thin disk components which would be spatially coincident with thin, mostly neutral mid-plane gas, and contribute most of the emission when viewed face-on. Moreover, recent observations of structures within our Galaxy have found synchrotron emission arising from cold, neutral gas as well \citep{Bracco2023}.
 
 While this is now known, the conventional wisdom when applying equipartition models to observations of extra-galactic non-thermal radio continuum emission has been to implicitly assume that the emission arises from a volume-filling phase of the ISM which is far more extended in the vertical direction compared to the atomic gas disk. This assumption can lead to underestimating the "true" \textbf{B} of the synchrotron-emitting dense gas, and over-estimating \textbf{B} in the volume-filling, tenuous thick disk/halo gas at $\sim$kpc above the disk.

We have found that explicitly evolving the CR(e) spectra is important for accurate synchrotron predictions towards galactic centers, where loss terms are drastically different from typical spiral galaxy conditions. Comparing to the "single-bin" scenario shows that the resulting predicted emission changes relative to assuming a constant LISM spectrum by modest factors of $\sim$1.5-2 in typical spiral galaxy conditions at outer radii (R $>$ 3 kpc), but can be particularly important by a factor $\sim$10 - 50 towards the galactic center, where loss terms can be drastically different. This is consistent with the softer CRe spectrum seen in these simulations towards the galactic center in \citet{hopkins_first_2022}, and highlights the varying importance of different loss rates and CR source distributions in generating predictive spectra and synchrotron images. We show that the difference in synchrotron owes primarily to variation in the p/e$^{-}$ ratio compounding with changes in spectral slope, rather than e$^{+}$ contributions.

Finally, we formulate a toy model that accounts for clumping factors, a varied magnetic scale height, and deviations from equipartition in order to more robustly trace \textbf{B} weighted by different quantities. From our toy model calculations, we find that uncertainty in connecting the equipartition values to the "true" values of u$_{\rm B}$ and e$_{\rm CR}$ boils down to the assumption of energy equipartition and the size/volume-filling factor of the emitting regions, and less so on spectral effects. When estimating a volume-averaged mean $\langle u_{\rm B} \rangle$, the fact that CRs are diffusive and so naturally form a smoother distribution, means that there are large local violations of equipartition. But this smoothness partially cancels the "clumping factor" from in-homogeneous \textbf{B}. This leads to surprisingly reasonable values of \textbf{B}. Future high-resolution radio observations at GHz frequencies \citep{Murphy2018} may further constrain the volume filling and clumping factors of synchrotron emission, including the current observationally uncertain thin-disk scale heights. 

We note that while we have only studied simulations with constant power-law scattering (or diffusivity) of CRs in this work, we have also studied a set of FIRE-2 simulations with varied CR transport motivated by extrinsic turbulence and self-confinement in \citet{Ponnada2023b}, though with the caveat that those are "single-bin" simulations. There, we find that the CR transport physics can drive differences in gas phase structure, morphology, and  non-thermal properties, leading to markedly different synchrotron emission particularly for the self-confinement models relative to extrinsic turbulence or constant diffusivity models. Correspondingly, there are implications for what phase of the ISM dominates the emission and the distribution of gas in the disk and inner CGM, which would affect what one would infer with equipartition assumptions. In particular, in runs with self-confinement motivated CR transport, the strong trapping of CRs in regions of high e$_{\rm CR}$ can lead to non-linear CR-driven, magnetized winds and thus result in a more volume-filling, diffuse phase of the ISM dominating the emission, more in-line with fiducial equipartition assumptions, though those simulated galaxies differ in detail from observed star-forming L$^{\ast}$ galaxies, as we discuss in \citet{Ponnada2023b}. In future work, we will continue to investigate how varied CR transport physics can act to vary synchrotron properties using fully cosmological, spectrally-resolved CR-MHD galaxy simulations.

Synchrotron emission has also been used to estimate physical properties of CRe like their transport length and diffusion coefficient (with an assumed streaming or advection speed), or the CRe scale height \citep{heesen_radio_2021,heesen_diffusion_2023}. In future work, we will explore these estimators using our simulations. 

\section*{Acknowledgements}

We wish to recognize and acknowledge the past and present Gabrielino-Tongva people and their Indigenous lands upon which this research was conducted. Additionally, we thank the staff at our institutes, without whose endless efforts this work would not be possible during the ongoing pandemic. We thank the anonymous referee for their helpful comments which improved the quality of this manuscript. We thank Marco Padovani for providing lookup tables for the synchrotron emissivity calculation. SP thanks Dr. Viviana Casasola for providing gas surface density data for the galaxies compared to in this study, and thanks Dr. Aritra Basu for providing FITS files of radio continuum observations for the same galaxies. Support for SP and PFH was provided by NSF Research Grants 1911233, 20009234, 2108318, NSF CAREER grant 1455342, NASA grants 80NSSC18K0562, HST-AR-15800. GVP acknowledges support by NASA through the NASA Hubble Fellowship grant  \#HST-HF2-51444.001-A awarded  by  the  Space Telescope Science  Institute,  which  is  operated  by  the Association of Universities for Research in Astronomy, Incorporated, under NASA contract NAS5-26555. CBH is supported by NSF grant AAG-1911233 and NASA grants HST-AR-15800, HST-AR-16633, and HST-GO-16703.  Numerical calculations were run on the Caltech compute cluster "Wheeler," allocation AST21010 supported by the NSF and TACC, and NASA HEC SMD-16-7592. The Flatiron Institute is supported by the Simons Foundation. CAFG was supported by NSF through grants AST-2108230  and CAREER award AST-1652522; by NASA through grants 17-ATP17-0067 and 21-ATP21-0036; by STScI through grant HST-GO-16730.016-A; by CXO through grant TM2-23005X; and by the Research Corporation for Science Advancement through a Cottrell Scholar Award. ISB was supported by the DuBridge Postdoctoral Fellowship at Caltech. DK was supported by NSF grant AST2108314. KS acknowledges support from the Black Hole Initiative at Harvard University, which is funded by grants from the John Templeton Foundation and the Gordon and Betty Moore Foundation. This work was supported by NSF grant AST-2109127.

\section*{Data Availability}

The data supporting the plots within this article are available on reasonable request to the corresponding author. A public version of the GIZMO code is available at \url{http://www.tapir.caltech.edu/~phopkins/Site/GIZMO.html}. FIRE-2 simulations are publicly available \citep{Wetzel2022} at \url{http://flathub.flatironinstitute.org/fire}, though simulations including the physics of MHD and cosmic rays like those analyzed in this study are not yet publicly available. Additional data, including initial conditions and derived data products, are available at \url{https://fire.northwestern.edu/data/}.

\bibliographystyle{mnras}
\bibliography{Synchrotron} 



\clearpage
\appendix

\renewcommand{\thefigure}{A\arabic{figure}}

\section{Auxiliary Figures}
\setcounter{figure}{0}
\begin{figure}
    \centering
    \includegraphics[width=0.45\textwidth]{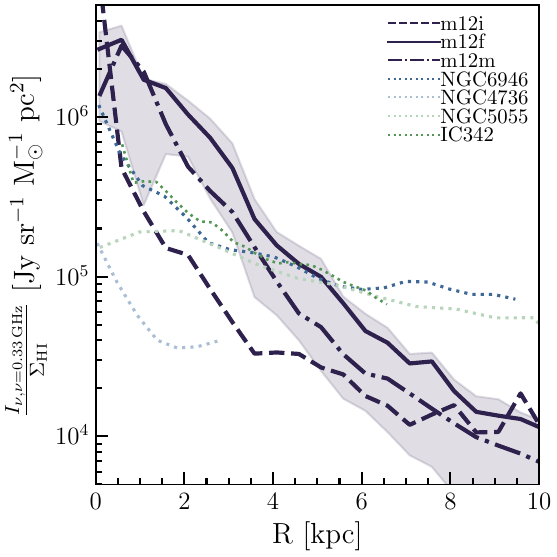}
    \caption{\textit{Azimuthally averaged radial profiles of synchrotron specific intensity normalized by HI surface density} for \texttt{m12i}, \texttt{m12f}, and \texttt{m12m}, in navy dashed, solid, and dot-dashed lines. Shaded regions show the approximate $\pm$ 1$\sigma$ scatter (32-68 percentile) at a given radial bin. Corresponding radial profiles for nearby spiral galaxies from \citet{basu_magnetic_2013} and \citet{beck_magnetic_2015} are shown in dotted lines, normalized by the $\Sigma_{HI}$ from \citep{casasola_radial_2017}. Normalized in this way, the radial profiles exhibit broader similarities when compared to the specific intensity profiles shown in Figure \ref{fig:rad_profiles}, though the simulations' radial profiles still remain steeper than the observations compared to here. }
    \label{fig:norm_profiles}
\end{figure}
\begin{figure*}
    \centering
    \includegraphics[width=1.0\textwidth]{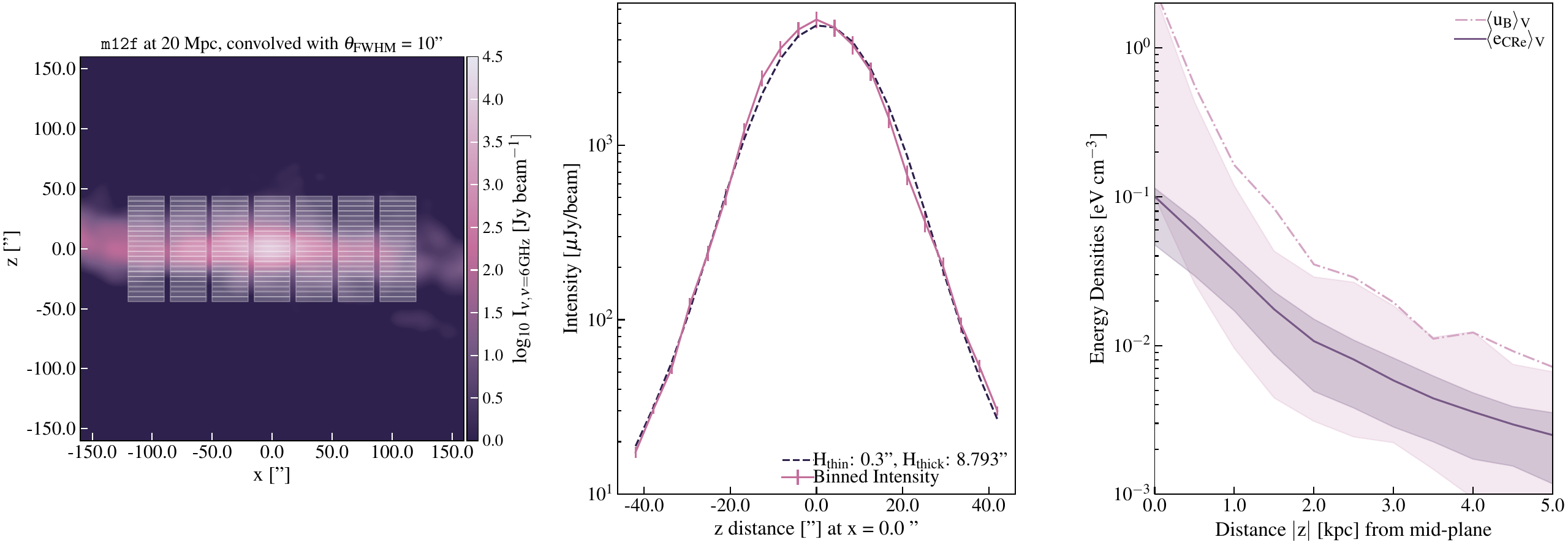}
    \caption{\textit{Edge-on image analysis:} \textbf{Left:} 10" Gaussian beam-convolved image of \texttt{m12f} at 6 GHz with the observer at a distance of 20 Mpc (1" $\sim$100 pc) with bins overlaid akin to the method of \citep{krause_chang-es_2018}. \textbf{Center}: Lines showing intensity profiles for the central strip of bins (pink solid) with best-fit two-component exponential fit overlaid (purple dashed). \textbf{Right:} Lines showing  vertical profiles of volume-weighted magnetic (pink dot-dashed) and CRe (navy solid) energy densities, with shaded regions showing 25-75 percentile ranges. Our results are consistent with observational constraints estimates of thin and thick disk components of edge-on spiral galaxies, with the thin disc component with small scale height ($\sim$120 pc when averaged across each vertical strip) dominating the emission and the thick disk component contributing little to the emission but extended out to scale heights of $\sim$kpc. This primarily owes to a small magnetic scale height, with u$_{\rm B}$ falling off more quickly than e$_{\rm CRe}$, which owing to CRe diffusion havs a flatter profile.}
    \label{fig:scale_heights}
\end{figure*}

\begin{figure}
    \centering
    \includegraphics[width=0.45\textwidth]{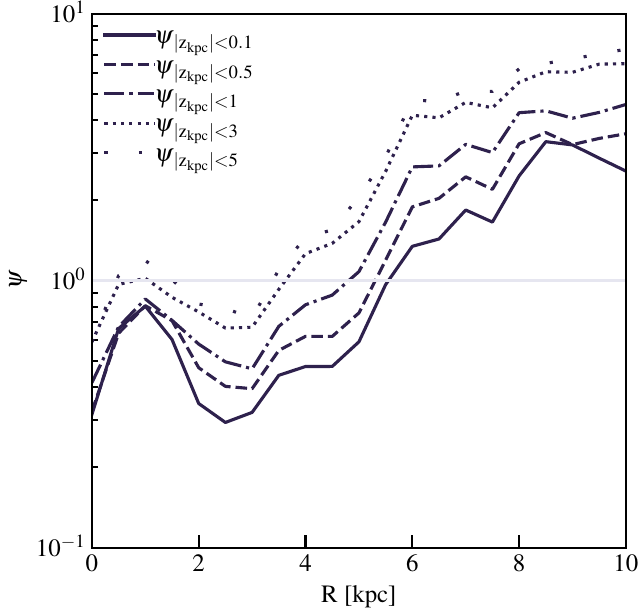}
    \caption{\textit{Radial profiles of $\psi$ (physical equipartition between \textbf{B} and CRs), or $\langle e_{\rm CR}\rangle_{\rm V}$/$\langle u_{\rm B}\rangle_{\rm V}$,} in cylindrical annuli of varying height from the mid-plane for \texttt{m12f}. For a cylindrical annulus of a given height, in the inner disk where gas densities are high, magnetic energy densities can dominate CR energy densities at the factor of $\sim$ 2 level for cylindrical heights $\leq$ 1 kpc in this volume-averaged sense, with this trend increasing as more of the dense mid-plane gas is sampled (lower heights). In the outer-disk, where gas densities are lower, CRs start to dominate the relative energy density, with the effect accentuated by sampling more "halo" gas at larger heights above the disk where e$_{\rm CR}$ tends to dominate u$_{\rm B}$ by factors of $\sim$ 2-4 for heights $\leq$ 1 kpc at the very outer radii \citep{Ponnada2022}.}
    \label{fig:psi_profile}
\end{figure}


\bsp	
\label{lastpage}
\end{document}